# Reading and Writing Single-Atom Magnets

Fabian D. Natterer[1, 2,*], Kai Yang[1, 3], William Paul[1], Philip Willke[1, 4], Taeyoung Choi[1], Thomas Greber[1, 5], Andreas J. Heinrich[1], and Christopher P. Lutz[1,*]

[1]IBM Almaden Research Center, San Jose, CA 95120, USA
[2]Institute of Condensed Matter Physics, École Polytechnique Fédérale de Lausanne, CH-1015 Lausanne, Switzerland
[3]School of Physical Sciences and Key Laboratory of Vacuum Physics, University of Chinese Academy of Sciences, Beijing 100049, China
[4]IV. Physical Institute, University of Göttingen, Friedrich-Hund-Platz 1, D-37077 Göttingen, Germany
[5]Physik-Institut, Universität Zürich, Winterthurerstrasse 190, CH-8057 Zürich, Switzerland

**The highest-density magnetic storage media will code data in single-atom bits. To date, the smallest individually addressable bistable magnetic bits on surfaces consist of 5-12 atoms[1,2]. Long magnetic relaxation times were demonstrated in molecular magnets containing one lanthanide atom[3–11], and recently in ensembles of single holmium (Ho) atoms supported on magnesium oxide (MgO)[12]. Those experiments indicated the possibility for data storage at the fundamental limit, but it remained unclear how to access the individual magnetic centers. Here we demonstrate the reading and writing of individual Ho atoms on MgO, and show that they independently retain their magnetic information over many hours. We read the Ho states by tunnel magnetoresistance[13,14] and write with current pulses using a scanning tunneling microscope. The magnetic origin of the long-lived states is confirmed by single-atom electron paramagnetic resonance (EPR)[15] on a nearby Fe sensor atom, which shows that Ho has a large out-of-plane moment of $(10.1 \pm 0.1)$ μ$_B$ on this surface. In order to demonstrate independent reading and writing, we built an atomic scale structure with two Ho bits to which we write the four possible states and which we read out remotely by EPR. The high magnetic stability combined with electrical reading and writing shows that single-atom magnetic memory is possible.**

The demonstration of magnetic bistability in single molecule magnets containing one rare earth atom[3,5,8,10] illustrated the potential of single-atom spin centers in future storage media[4,6,7,9,11]. A ligand field that provides a barrier against magnetization reversal by lifting the Hund degeneracies in single molecule magnets[3–10] can also be realized for atoms bound to a surface[16–19]. While a break junction probes the quantum states of one isolated molecule[20], a surface enables preparation of and access to numerous spin centers. Magnetic lifetimes in the milliseconds range were accordingly obtained for single 3d atoms on magnesium oxide (MgO)[21] but the report of magnetic bistability for Ho atoms on a platinum surface is debated[22–25]. A major advance of observing magnetic remanence was recently achieved with an ensemble of isolated Ho atoms on MgO[12], yet the question remained whether electrical probing of the highly localized f orbitals of individual rare earth atoms is possible[23,26,27].

Here we address the magnetic switching of individual Ho atoms on MgO, which we control by current pulses and detect by changes to the tunnel magnetoresistance using a spin-polarized scanning tunneling microscope (STM)[13]. We prove the magnetic origin of the switching in the tunneling resistance by STM-enabled single-atom electron paramagnetic resonance (EPR) on an adjacent iron (Fe) sensor atom. Additionally, we determine by this method the out-of-plane component of the Ho magnetic moment, and use the long lifetime to store two bits of information in an array of two Ho atoms whose magnetic state can be read remotely by means of EPR on a nearby sensor atom.

Figure 1 shows our experimental setup consisting of a low-temperature STM with EPR capabilities[15]. The latter feature allows the use of Fe atoms as local magnetometers to determine nearby magnetic moments[28]. Upon dosing Ho on bilayer MgO, we find individually adsorbed Ho atoms in two binding sites: atop oxygen and on bridge sites. In the present work we will focus on Ho atoms atop oxygen – the Ho species showing long lifetime[12] – and note that we can move individual Ho atoms from bridge to oxygen sites by using voltage pulses. The co-adsorbed Fe sensor atoms can be distinguished from Ho by a lower topographic height and by their spectroscopic fingerprint: Fe atoms show inelastic spin excitations at ~14 meV[19]. Holmium atoms, on the other hand, are devoid of spin-excitation signatures. However, we measure a two-state signal on Ho atoms showing discrete changes in conductance of up to 4% with a spin-polarized tip (Figure 1**b**). The current trace has plateaus of long residence times in the high and low-conductance states. For the bias voltage $V = 150$ mV employed in Figure 1**c**, the magnetic residence time ranges from dozens of seconds to fractions of a second for tunnel currents $I$ between 6 pA and 600 pA. The linear increase of the switching rate with tunnel current indicates a single-electron rate-limiting process having a miniscule switching probability of order $10^{-9}$ per tunneling electron. We observe that the Ho states remain stable for hours when the bias voltage is kept below $|V| \approx 65$ mV. At higher voltages we see an increasing switching rate of the Ho states with bias voltage (Figure 1**d**). We therefore have two means to control the switching rate: the tunnel current as well as the tunnel voltage. To write the Ho state, we subject the Ho atom repeatedly to a current pulse at $V > 150$ mV, until we detect a change of magnetoresistance at $V = 50$ mV that indicates the Ho is in the desired final state. Note that three rate-increasing voltage thresholds appear in Figure 1**d.** These thresholds reflect transition energies between different magnetic states of Ho on MgO.

In the following, we will use a nearby Fe sensor atom to show that the two Ho states correspond to two magnetic orientations of the Ho moment. The Fe sensor acts as a local magnetometer[28], since the Zeeman splitting of its ground state responds to the dipolar field of the nearby Ho atom. The Zeeman splitting of the Fe sensor, dominated by the external out-of-plane magnetic field $B_z$, is therefore shifted to lower frequency when the Ho moment is aligned in the direction of $B_z$, and to higher frequency when it is aligned opposite to $B_z$. The EPR spectrum on the Fe sensor in Figure 2**a** revealed a single resonance peak when Ho was in its high conductance state. After switching Ho to the low conductance state, the EPR peak correspondingly shifted to lower frequency. We find that the frequency difference ($\Delta f$) sensitively depends on the Fe-Ho distance, as seen in Figure 2**b**. Following previous work[28], we describe the $\Delta f$ vs distance scaling in terms of the magnetic dipole-dipole interaction for out-of-plane polarized moments. Using the Fe moment of $(5.44 \pm 0.03)$ $\mu_B$ on the same MgO/Ag(100) surface[28], a one-parameter fit yields the Ho moment of $(10.1 \pm 0.1)$ $\mu_B$. The observed Ho magnetic moment of ~10 $\mu_B$ suggests a $4f^{10}$ Ho(III) ion configuration where its total angular momentum $J$ is polarized out-of-plane ($J_z = \pm 8$). This exceeds the $J_z$ value deduced from ensemble measurements by Donati and co-workers[12] and is possibly related to the averaging over different adsorption sites for Ho on MgO in the ensemble measurements. Most importantly, however, the correlation of Ho in high and low conductance states to the respective EPR frequencies unambiguously proves the magnetic origin of the two-state switching for Ho. In addition, the presence of only one EPR peak in each spectrum acquired over many minutes is a consequence of the extraordinary stability of the magnetic state of the Ho atom.

The high stability of the Ho moment could find use in single-atom data storage applications. To exemplify this point, we built a stable two-bit atomic Ho array (Figure 3**a**) and measured the Ho states non-invasively by EPR on a nearby Fe sensor. The frequencies of the four EPR lines can be predicted from our dipole-dipole data in Figure 2**b**. Accordingly, after setting the Ho bits individually to their up and down states, we observe one of four distinct EPR lines on the Fe sensor, identifying the four possible states of the Ho bit array. While writing the magnetic state of either of the atoms, we did not observe unintended magnetic switching of its neighbor.

The use of single-atom magnets as building blocks in complex structures should enable the study of magnetic interactions at the threshold between single-atom magnetism and collective behaviors emerging from dense magnetic arrays.

Electrical reading and writing of magnetic information for Ho at the single-atom limit offers new input into the miniaturization of storage media and renders transport measurement of the elusive f orbitals appealing for investigation of the magnetic properties of the rare earth atoms.

*Methods:* We performed the experiments in a home built Joule-Thomson low-temperature STM, working in ultra-high vacuum at 1.2 K, and carried out EPR measurements by adding a radio frequency voltage of typically $V_{RF}$ = 15 mV amplitude at 10–30 GHz onto the DC bias voltage[15,29]. A magnetic field of 0.9 T was applied with a $B_z$ component of ~100 mT providing a Zeeman splitting of ~20.8 GHz for our Fe sensors. We dosed Ho and Fe atoms from pieces of pure metal with e-beam evaporators directly onto the sample held at a temperature below 10 K. The STM tip was an Ir wire that had been conditioned by field emission and subsequent indentations into the Ag crystal until it became coated by Ag and atomically sharp. We furthermore spin polarized our tips by picking up Fe atoms (typically one to five) until pump-probe measurements on Fe sensors yielded spin contrast. The Ho switching rate in Figures 1**c** and **d** accounts for the total number of switches; from up to down state and vice versa. The coefficients of the piecewise linear fit $\Gamma = \sum_{i=1}^{3} c_i (V - V_i)/V_i$ in Figure 1**d** are $c_1 = (0.20 \pm 0.01)$, $c_2 = (14.0 \pm 0.6)$, and $c_3 = (89 \pm 3)$ s$^{-1}$. To grow MgO, we exposed an atomically clean Ag(100) single crystal, held at ~600 K, to an Mg flux from a Knudsen cell in an oxygen partial pressure of ~10$^{-6}$ mbar up to a MgO coverage of 1.5 monolayers. The growth rate was ~0.5 monolayers per minute[30,31]. We assumed in-plane commensurability between the Ag and MgO lattices and a low-temperature Ag lattice parameter of 408.67 pm[28]. All atoms in this paper were studied on bilayer MgO as determined by the conductance at point contact[21]. Note that this bilayer thickness had been previously denoted as monolayer MgO[15].

**Acknowledgement**

We thank Bruce Melior for expert technical assistance and F. Donati for fruitful discussions. We gratefully acknowledge financial support from the Office of Naval Research. F.D.N. greatly appreciates support from the Swiss National Science Foundation under project numbers P300P2_158468 and PZ00P2_167965. K.Y. acknowledges support from National Natural Science Foundation of China (Grant No. 61471337). W.P. thanks the Natural Sciences and Engineering Research Council of Canada for fellowship support. P.W. gratefully acknowledges financial support from the German academic exchange service. T.G. thanks IBM Research for its hospitality.

**Figure Captions**

**Figure 1 | Experimental setup and magnetic switching of Holmium. a**, Topographic image of a Ho and Fe atom on bilayer MgO. The magnetic states of Ho are controlled and probed with an STM using spin-polarized tunneling ($V = 10$ mV, $I = 10$ pA, $2.35 \times 2.35$ nm$^2$, $T = 1.2$ K). **b**, The magnetoresistive tunnel current recorded atop a Ho atom shows switching between two magnetic states of long residence time ($V = 150$ mV, $I = 25$ pA). At these tunneling conditions, switching is induced by tunneling electrons. **c**, The switching rate scales as $\Gamma = a \times (I/I_0)^N$ where N = $0.95 \pm 0.01$ and $a = (1.5 \pm 0.1) \times 10^{-2}$ s$^{-1}$ ($V = 150$ mV, $I_0 = 1$ pA). **d**, The voltage dependence of $\Gamma$ at constant current reveals three rate-increasing thresholds at $V_1 = (73 \pm 1)$, $V_2 = (104 \pm 1)$, and $V_3 = (119 \pm 1)$ mV. The solid line is a three segment piecewise linear fit. The error bars of the fits in **c** and **d** indicate the standard error on the least-squares fit parameter.

**Figure 2 | Controlling the magnetic states of Holmium. a**, EPR spectrum of an Fe sensor for Ho in the high conductance (top) and low conductance (bottom) state. The continuous black lines are fits to a single Lorentzian and were used to extract the resonance frequencies of the EPR signal. ($V = 60$ mV, $I = 20$ pA, $V_{RF} = 15$ mV). **b**, The plot of $\Delta f$ for different Fe-Ho distances shows an excellent agreement with a magnetic dipole-dipole interaction and follows a power law of negative slope 3. The one parameter fit yields a Ho magnetic moment of $(10.1 \pm 0.1)$ μ$_B$. The uncertainty accounts for the standard error on the least-squares fit parameter and the error propagation from the Fe sensor magnetic moment. Measurements of different Ho atoms (coded with distinct symbols) show identical moments.

**Figure 3 | Example of a stable two-bit atomic Ho array. a**, The Ho atoms were arranged in a (x,y) = (4,4) and (8,3) configuration with respect to the Fe sensor, measured in increments of the oxygen sub-lattice ($V = 60$ mV, $I = 6$ pA). **b**, Normalized EPR spectra demonstrating simultaneous readout as measured on the Fe sensor atom. All four states can be read out in a single EPR spectrum. The only change in the magnetic states occurred after deliberately switching the Ho atoms and no spontaneous reversal was observed over several hours. The black solid lines are fits to Lorentzian curves. ($V = 60$ mV, $I = 20$ pA, $V_{RF} = 15$ mV).

**Figure 1**

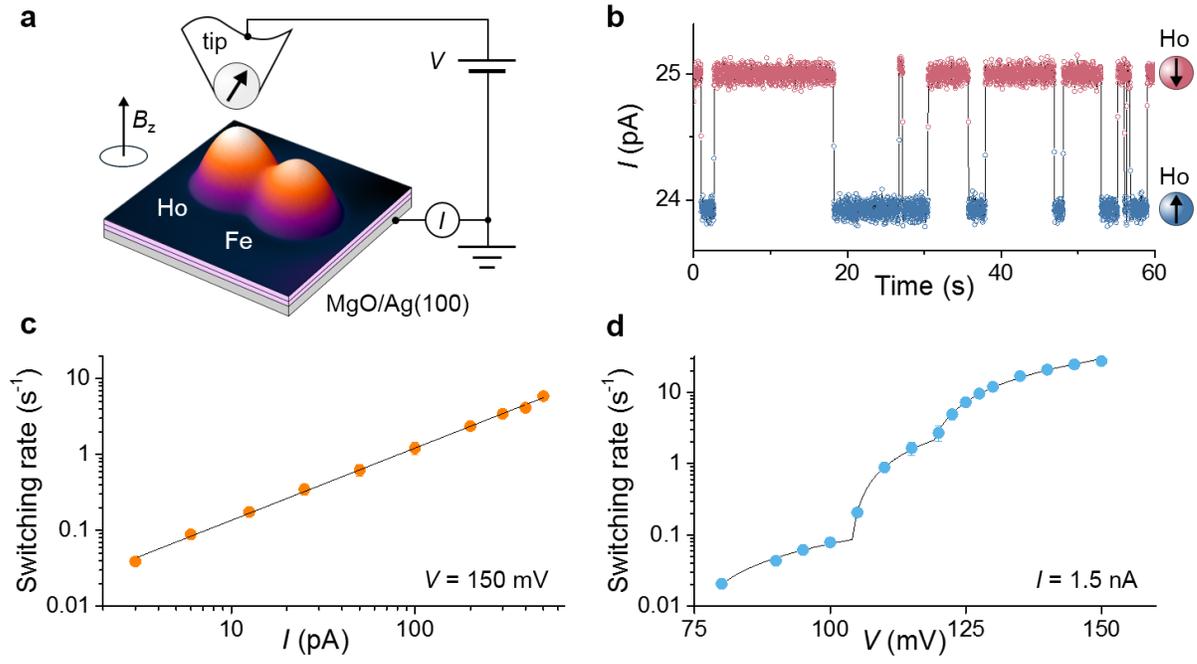

**Figure 2**

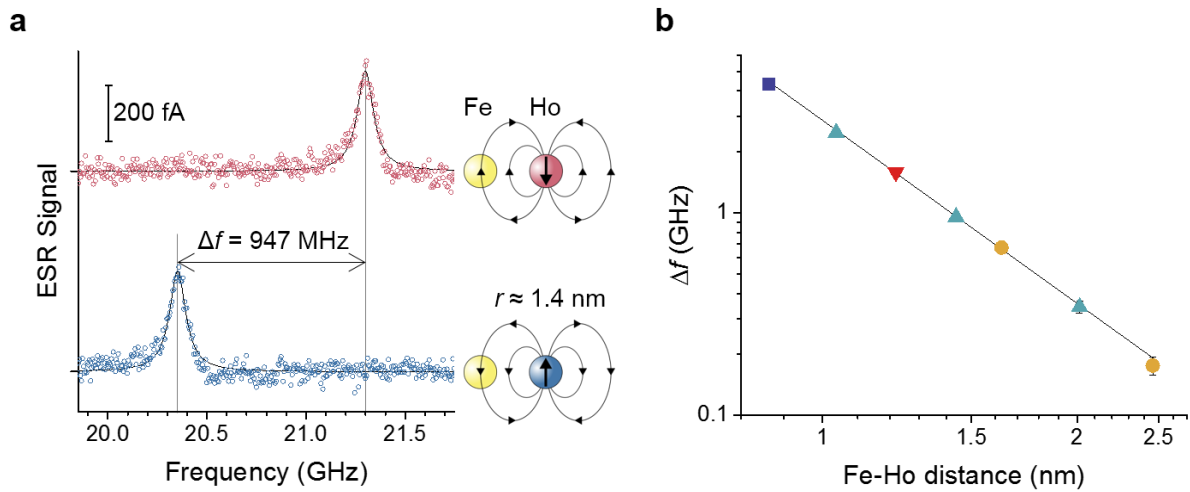

**Figure 3**

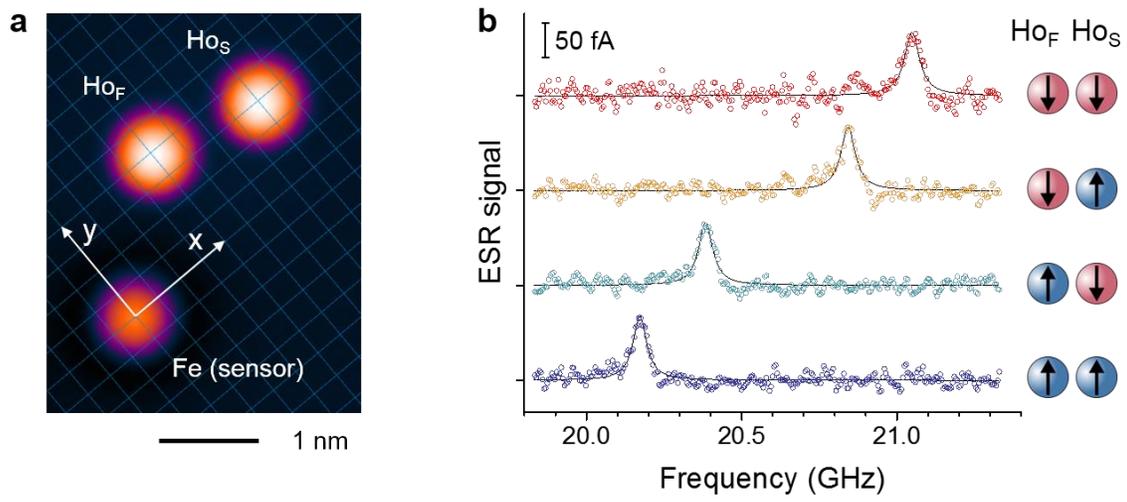

# References

\* To whom all correspondence should be addressed: donat.natterer@gmail.com or cplutz@us.ibm.com